\documentclass[nofootinbib,aps,prstab,reprint,groupedaddress,superscriptaddress,showpacs]{revtex4-2}
\usepackage[english]{babel}
\usepackage[utf8x]{inputenc}
\usepackage{amssymb}
\usepackage{amsthm}
\usepackage{amsmath}
\usepackage{hyperref}
\usepackage{subfigure}
\usepackage[percent]{overpic} 
\usepackage{picture}
\usepackage{color,soul}
\usepackage{nth}
\usepackage{gensymb}
\usepackage{cleveref}
\usepackage{textcomp}
\usepackage{makecell}
\usepackage{lineno} 
\usepackage{textcomp}

\graphicspath{{img/}}

\begin{document}

\title{Recovery of Hydrogen plasma at the sub-nanosecond timescale in a plasma-wakefield accelerator}

\author{R.~Pompili}
\email[]{riccardo.pompili@lnf.infn.it}
\affiliation{Laboratori Nazionali di Frascati, Via Enrico Fermi 54, 00044 Frascati, Italy}
\author{M.P.~Anania}
\author{A.~Biagioni}
\affiliation{Laboratori Nazionali di Frascati, Via Enrico Fermi 54, 00044 Frascati, Italy}
\author{M.~Carillo}
\author{E.~Chiadroni}
\affiliation{University of Rome Sapienza, Piazzale Aldo Moro 5, 00185 Rome, Italy}
\author{A.~Cianchi}
\affiliation{University of Rome Tor Vergata, Via della Ricerca Scientifica 1, 00133 Rome, Italy}
\affiliation{INFN Tor Vergata, Via della Ricerca Scientifica 1, 00133 Rome, Italy}
\affiliation{NAST Center, Via della Ricerca Scientifica 1, 00133 Rome, Italy}
\author{G.~Costa}
\author{L.~Crincoli}
\author{A.~Del~Dotto}
\author{M.~Del~Giorno}
\affiliation{Laboratori Nazionali di Frascati, Via Enrico Fermi 54, 00044 Frascati, Italy}
\author{F.~Demurtas}
\affiliation{University of Rome Tor Vergata, Via della Ricerca Scientifica 1, 00133 Rome, Italy}
\author{M.~Ferrario}
\affiliation{Laboratori Nazionali di Frascati, Via Enrico Fermi 54, 00044 Frascati, Italy}
\author{M.~Galletti}
\affiliation{University of Rome Tor Vergata, Via della Ricerca Scientifica 1, 00133 Rome, Italy}
\affiliation{INFN Tor Vergata, Via della Ricerca Scientifica 1, 00133 Rome, Italy}
\affiliation{NAST Center, Via della Ricerca Scientifica 1, 00133 Rome, Italy}
\author{A.~Giribono}
\affiliation{Laboratori Nazionali di Frascati, Via Enrico Fermi 54, 00044 Frascati, Italy}
\author{J.K.~Jones}
\affiliation{ASTeC, STFC Daresbury Laboratory, Sci-Tech Daresbury, WA4 4AD, Warrington, UK}
\author{V.~Lollo}
\affiliation{Laboratori Nazionali di Frascati, Via Enrico Fermi 54, 00044 Frascati, Italy}
\author{T.~Pacey}
\affiliation{ASTeC, STFC Daresbury Laboratory, Sci-Tech Daresbury, WA4 4AD, Warrington, UK}
\author{G.~Parise}
\affiliation{University of Rome Tor Vergata, Via della Ricerca Scientifica 1, 00133 Rome, Italy}
\author{G.~Di~Pirro}
\author{S.~Romeo}
\affiliation{Laboratori Nazionali di Frascati, Via Enrico Fermi 54, 00044 Frascati, Italy}
\author{G.J.~Silvi}
\affiliation{University of Rome Sapienza, Piazzale Aldo Moro 5, 00185 Rome, Italy}
\author{V.~Shpakov}
\author{F.~Villa}
\affiliation{Laboratori Nazionali di Frascati, Via Enrico Fermi 54, 00044 Frascati, Italy}
\author{A.~Zigler}
\affiliation{Racah Institute of Physics, Hebrew University, 91904 Jerusalem, Israel}


\date{\today}


\begin{abstract}
Plasma wakefield acceleration revolutionized the field of particle accelerators by generating gigavolt-per-centimeter fields. 
To compete with conventional radio-frequency (RF) accelerators, plasma technology must demonstrate operation at high repetition rates, with a recent research showing feasibility at megahertz levels using an Argon source that recovered after about 60~ns. 
Here we report about a proof-of-principle experiment that demonstrates the recovery of a Hydrogen plasma at the sub-nanosecond timescale.
The result is obtained with a pump-and-probe setup and has been characterized for a wide range of plasma densities. We observed that large plasma densities reestablish their initial state soon after the injection of the pump beam ($<0.7$~ns). Conversely, at lower densities we observe the formation of a local dense plasma channel affecting the probe beam dynamics even at long delay times ($>13$~ns). 
The results are supported with numerical simulations and represent a step forward for the next-generation of compact high-repetition rate accelerators. 
\end{abstract}



\maketitle

\section*{Introduction}
Particle accelerators enabled a groundbreaking perspective of matter at the sub-atomic level~\cite{yoneda2015atomic} and sub-femtosecond timescale~\cite{maroju2020attosecond}.
Nowadays accelerator technology is based on RF which is ultimately limited in terms of achievable gradients (and, thus, sizeable footprints) by electrical breakdown~\cite{argyropoulos2018design}.
Plasma acceleration represented a breakthrough~\cite{tajima_dawson}, with experiments that demonstrated accelerations of  gigaelectronvolt per meter over short distances~\cite{2002Sci...298.1596M,2007Natur.445..741B,mangles2004monoenergetic,faure2004laser,litos2014high,adli2018acceleration,gonsalves2019petawatt} of high-quality beams~\cite{deng2019generation,pompili2021energy,pompili2022free}.
However, the generation of large fields in plasma triggers the motion of plasma ions~\cite{gorbunov2001plasma,rosenzweig2005effects} making necessary to wait for the recovery of the plasma to its unperturbed state. This ultimately defines the largest sustainable repetition rate for a plasma accelerator, with a recent work that has shown the feasibility to reach megahertz rates with an Argon plasma that recovered after about $60$~ns~\cite{d2022recovery}.


\begin{figure}[!b]
\centering
\includegraphics[width=1.0\linewidth]{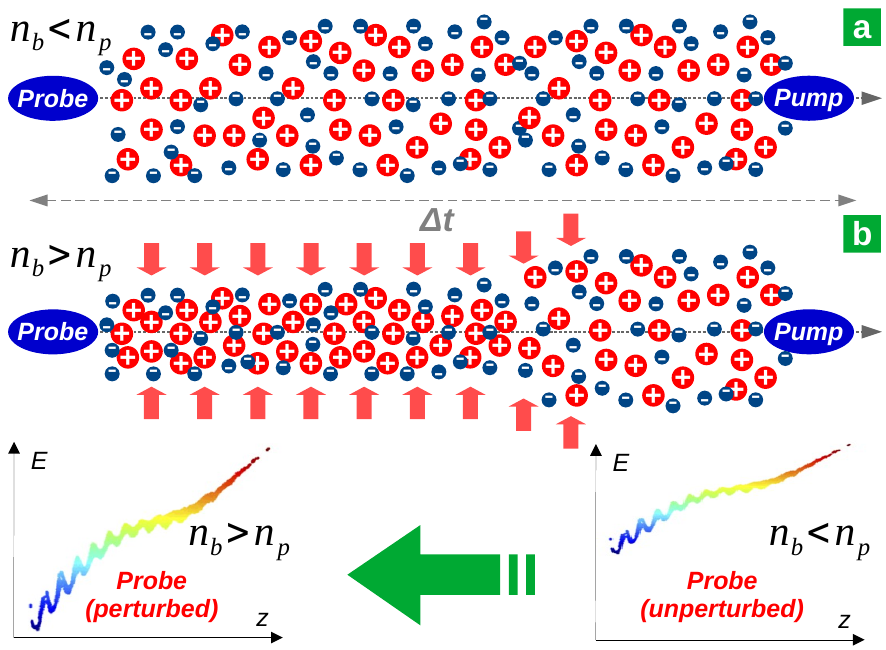}
\caption{Sketch of the experiment. The pump electron beam moves in the plasma undergoing deceleration. Depending on the plasma ($n_p$) and beam ($n_b$) densities, the ions stay mostly unperturbed (a) or are pinched (b, generating larger $n_p$). This affects the probe beam (delayed by $\Delta t$ with respect to the pump) whose deceleration is proportional to $n_p$. The two plasma-modulated longitudinal phase-spaces highlight the different signatures on the probe.}
\label{PinchSetup}
\end{figure}

\begin{figure*}[t]
\centering
\begin{overpic}[width=0.99\linewidth]{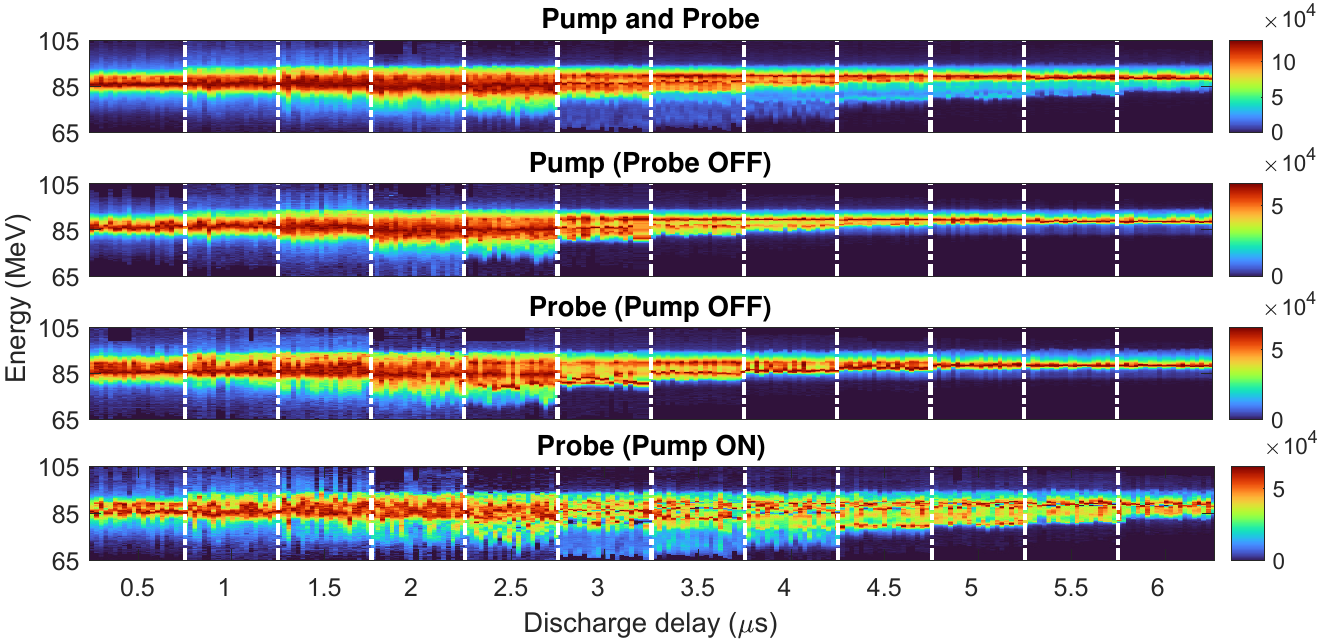} 
\put(90,39){\color{white}\textbf{a}}
\put(90,28){\color{white}\textbf{b}}
\put(90,17){\color{white}\textbf{c}}
\put(90,6.5){\color{white}\textbf{d}}
\end{overpic}
\caption{Experimental measurements with pump-probe delay $\Delta t\approx 0.7$~ns. The plots are obtained by stacking 20 consecutive shots (delimited by the dashed white lines) acquired in the range $\tau_D=0.5-6~\mu s$ with steps of $0.5~\mu s$. (a-d) Energy spectra collected at several $\tau_D$ (corresponding to different plasma densities) with both pump and probe beams (a), only pump (probe off, b) and only probe (pump off, c). The plot (d) shows the reconstructed probe (with pump on) obtained as $d=a-b$.}
\label{dPlot_13ns}
\end{figure*}
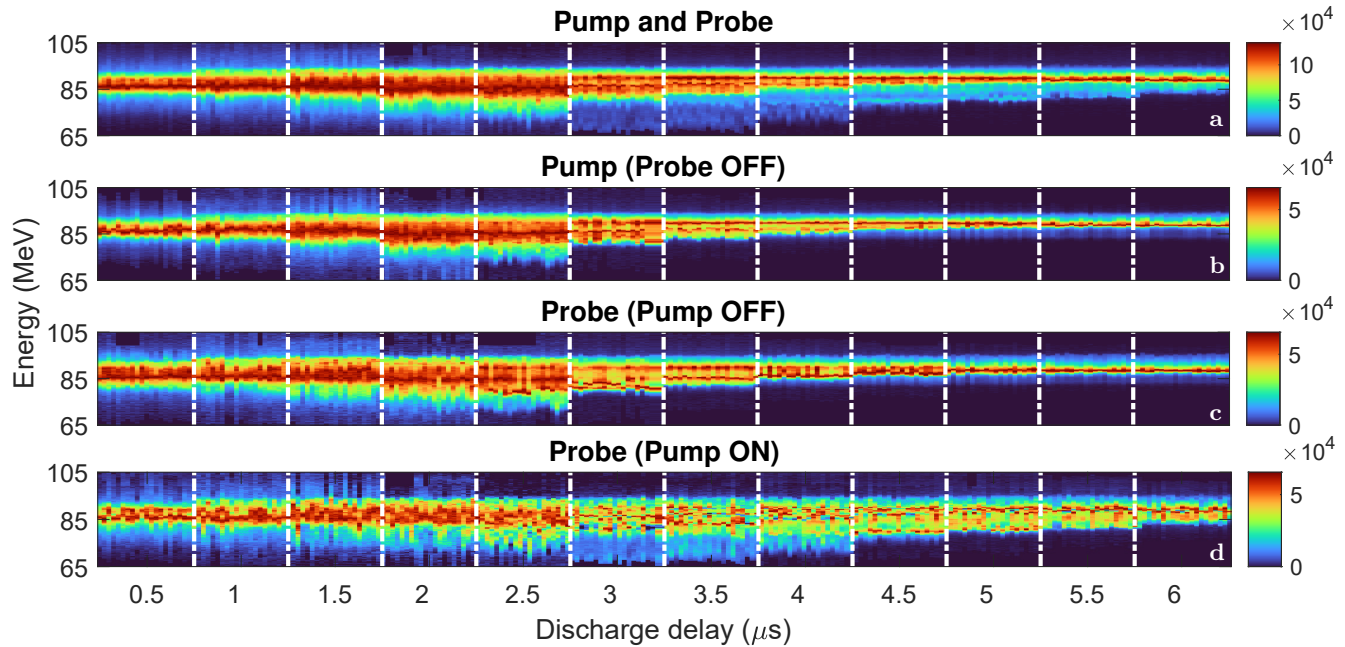

Here we report about a proof-of-principle experiment demonstrating sub-nanosecond recovery of a Hydrogen plasma, potentially enabling the operation of future particle accelerators at even larger repetition rates.
The experiment, depicted in Fig.~\ref{PinchSetup}, employs a pump-and-probe configuration with the first electron beam exciting the plasma wakefield and the second one probing the perturbation introduced in the plasma.
Since each beam (with density $n_b\approx 2\times 10^{15}$~cm$^{-3}$) generates a decelerating wakefield that depends on the plasma density $n_p$, by measuring the probe deceleration we retrieved the perturbation induced in the plasma by the pump. 
To provide a comprehensive picture of the interaction process, we performed a parametric study by exploring a wide range of plasma densities ($n_p\approx 10^{13}- 10^{16}$~cm$^{-3}$) and pump-probe delays ($\Delta t=0.7-13$~ns).
Defining $\alpha=n_b/n_p$ as the density ratio~\cite{geraci2000transverse,kumar2010self}, the results highlight (i) a negligible effect induced by the pump when $\alpha<1$ and (ii) the generation of a denser plasma channel when $\alpha>1$.

\section*{Results}
The experiment has been performed at the SPARC\_LAB test-facility~\cite{ferrario2013sparc_lab} by employing electron beams with 500~pC charge, $84\pm0.1$~MeV energy, $167\pm 14$~fs duration and $\sigma_r=43\pm 2~\mu m$ spot size at the plasma entrance measured with a transition radiation (TR) screen installed below the capillary in correspondence of its entrance. The errors are computed as the standard deviations. 
The two bunches are generated by the SPARC\_LAB photo-injector, consisting of a RF gun followed by three accelerating sections. The bunches come from the photo-emission of a copper cathode illuminated by two ultraviolet pulses~\cite{2011NIMPA.637S..43F,pompili2021time} and are temporally compressed with velocity-bunching~\cite{pompili2016beam}. To reach tens of nanoseconds delays between the bunches, the probe laser is generated by sending a replica of the pump laser through a several-meter long delay line. 
The plasma, confined in a 3~cm-long sapphire capillary with 2~mm hole diameter, is generated by ionizing Hydrogen gas with a high-voltage discharge (7~kV and 230~A). Two triplets of movable permanent magnet quadrupoles (PMQs) are installed upstream and downstream of the capillary to focus the beam into the plasma and at the extraction from it after the interaction~\cite{pompili2018compact}.

\begin{figure}[b]
\centering
\includegraphics[width=1.0\linewidth]{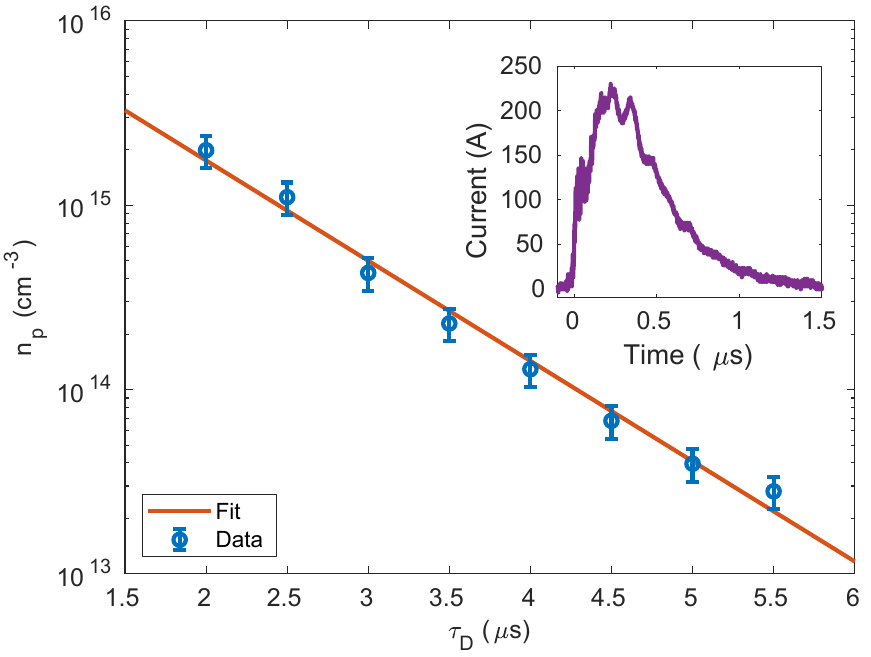}
\caption{Estimation of plasma density. The average plasma density is probed with the electron beam for several discharge delays $\tau_D$ (blue points). The red line reports the corresponding exponential fit $n_p\propto \exp{\left({-\tau_D/\tau_R}\right)}$ with $\tau_R\approx 0.7~\mu s$. The inset shows the discharge current waveform. The errorbars are computed as the standard deviations of 50 consecutive shots.}
\label{TimeDensityFit}
\end{figure}

\begin{figure*}[t]
\centering
\subfigure{
\begin{overpic}[height=0.36\linewidth]{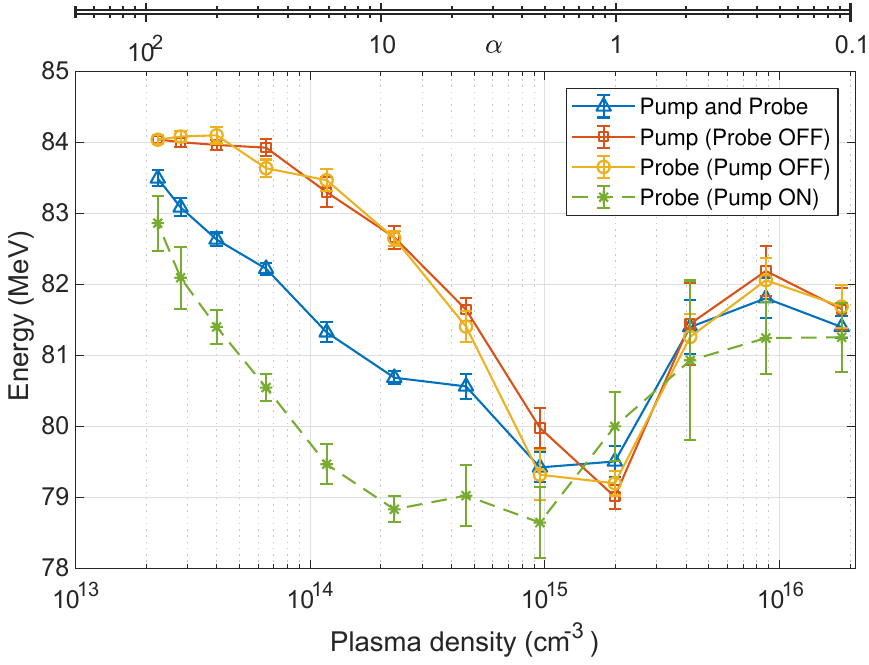}
\put(11,63){\color{black}\textbf{a}}
\end{overpic}
\label{Energy_13ns}
}
\subfigure{
\begin{overpic}[height=0.36\linewidth]{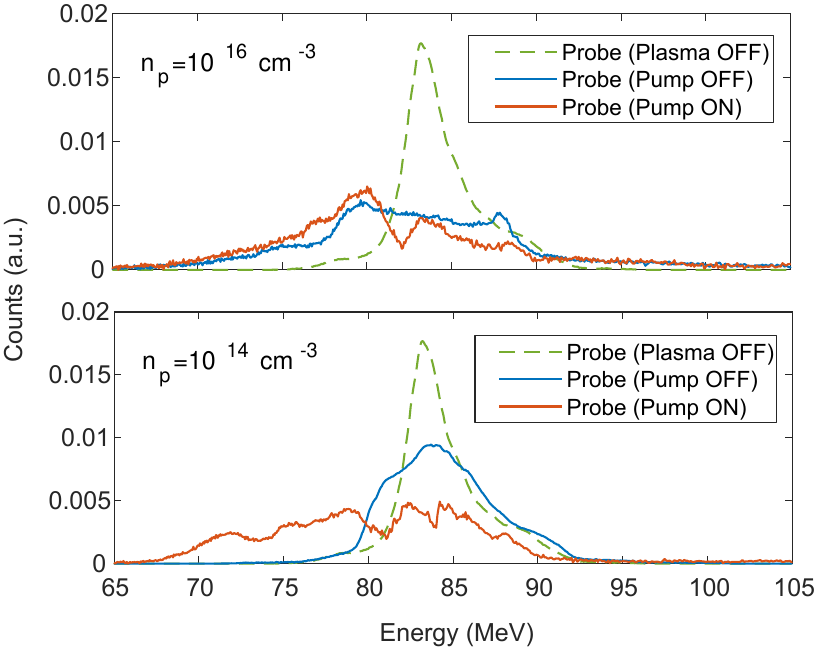}
\put(91,22){\color{black}\textbf{b}}
\end{overpic}
\label{eSpectrum_13ns}
}
\caption{Data with pump-probe delay $\Delta t=0.7$~ns. (a) Mean energies for the spectra containing both the pump and probe (blue) and the two alone (red and yellow). The dashed green line shows the mean energies of the perturbed probe, obtained by subtracting to the pump and probe spectra the ones containing only the pump.  The errorbars are computed as the standard deviations of 20 consecutive shots. (b) Single-shot probe energy spectra with (red) and without (blue) the pump at initial plasma density $n_p\approx 10^{16}$~cm$^{-3}$ ($\tau_D\approx 1~\mu s$, top) and $n_p\approx 10^{14}$~cm$^{-3}$ ($\tau_D\approx 4~\mu s$, bottom). The dashed green line shows the probe spectrum with plasma turned off.}
\label{data_13ns}
\end{figure*}

Figure~\ref{dPlot_13ns} shows several energy spectra measured with the pump and probe beams sent together (delayed by $\Delta t=0.7$~ns) and one at a time. The spectra are measured for different plasma densities $n_p$, obtained by putting a delay $\tau_D$ on the beam arrival time with respect to the discharge current pulse.
This allows the tuning of $n_p$ by adjusting $\tau_D$ as shown in Fig.~\ref{TimeDensityFit}. This calibration curve highlights that, due to plasma recombination, larger $\tau_D$ corresponds to smaller plasma densities~\cite{stefano_density} with an exponential decay time equal to $\tau_R\approx 0.7~\mu s$.
The energy spectra showed in Fig.~\ref{dPlot_13ns} are measured on the Cerium-doped Yttrium Aluminum Garnet (Ce:YAG) screen of a magnetic spectrometer.
The energy spectra of the perturbed probe (with pump on), in particular, are obtained by subtracting the pump beam traces from those containing both pump and probe beams. This highlights an enhanced deceleration at low densities ($\tau_D \gtrsim 2.5~\mu s$, corresponding to $n_p \lesssim 10^{15}$~cm$^{-3}$). On the contrary, at larger $n_p$ ($\tau_D\lesssim 2.5~\mu s$) the difference between the perturbed and unperturbed probe is negligible.

To quantify the effect we computed, at the same delay $\Delta t$, the resulting mean energies for all the beam configurations. Figure~\ref{Energy_13ns} shows that the single pump and probe bunches experienced the largest deceleration for $\alpha\approx 1$, i.e. at $n_p\approx 2\times 10^{15}$~cm$^{-3}$. The two bunches have approximately the same interaction with the plasma when sent one at a time, with the small differences arising from the slightly different dynamics they experienced in the photo-injector. Conversely, their dynamics in the plasma become very dissimilar when they are sent together.
We can see that there is a strong difference between the perturbed (pump on) and unperturbed (pump off) probe beams for low plasma densities in the range $n_p \approx 10^{13}-10^{15}$~cm$^{-3}$, i.e. $\alpha\approx 1-100$. On the contrary, for larger $n_p$ (with $\alpha<1$) the differences become negligible, indicating that the plasma recovered to its initial state in less than 0.7~ns.
This is emphasized in Fig.~\ref{eSpectrum_13ns}, that compares two energy spectra profiles of the perturbed and unperturbed probe at two selected $n_p$. One can see that for large plasma densities ($n_p \approx 10^{16}$~cm$^{-3}$, $\alpha<1$) there is no clear perturbation on the probe beam, i.e. in both cases the beam undergoes the same amount of deceleration, while for smaller plasma densities ($n_p \approx 10^{14}$~cm$^{-3}$, $\alpha>1$) the perturbed probe has a very different energy spectrum with respect to the unperturbed one.
A larger deceleration is indeed evident when the pump is turned on, with the probe energy spectrum showing tails at lower energies (down to $\approx 65$~MeV). This suggests that the probe interacted with a larger plasma density.
The plot also shows the energy spectrum of the probe beam when the discharge (and, in turn, the plasma) is turned off. No field ionization features~\cite{o2006plasma} are evident on the profile when the Hydrogen gas flows into the capillary, indicating that both the pump and probe beams did not ionize the gas.

\begin{figure}[ht]
\centering
\includegraphics[width=1.0\linewidth]{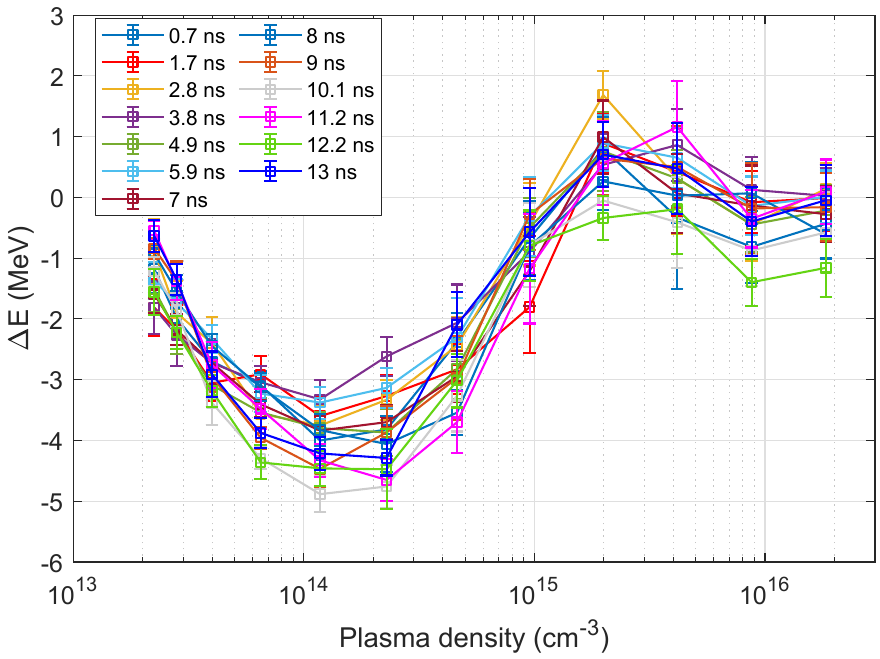}
\caption{Scan of plasma density $n_p$ and pump-probe delay $\Delta t$. The \textit{x} axis shows the plasma densities (corresponding to different discharge delays $\tau_D$) while \textit{y} axis reports the energy variation $\Delta E$ of the perturbed probe (pump on) with respect to the unperturbed one (pump off) for several delays $\Delta t$. Each point is obtained by averaging 20 consecutive shots. The errorbars are computed as the standard deviations these shots.}
\label{EnergyDiffAll}
\end{figure}

The data analysis has been repeated for several pump-probe delays in the range $\Delta t=0.7-13$~ns.
Figure~\ref{EnergyDiffAll} shows the energy difference $\Delta E$ between the unperturbed (pump off) and perturbed (pump on) probe as a function of the plasma density. Each line is obtained for a specific $\Delta t$ with each point given by the average of 20 consecutive shots. The plot indicates that the probe interaction with the plasma is similar for each $\Delta t$ and for a given value of $n_p$. The plasma rapidly recovers ($\Delta E\approx 0$) for $n_p\gtrsim 2\times 10^{15}$~cm$^{-3}$ ($\alpha<1$). On the contrary, at lower $n_p$ ($\alpha>1$) the perturbation induced by the pump is always present even after 13~ns ($\Delta E\neq 0$).
The largest difference in terms of $\Delta E$ is at $n_p\approx 10^{14}$~cm$^{-3}$, indicating that a strong perturbation is induced in the plasma when the pump is turned on. By comparing the perturbed probe deceleration in Fig.~\ref{Energy_13ns} we see that, for such a specific $n_p$, a similar deceleration of the unperturbed probe was reached at $n_p\approx 2\times 10^{15}$~cm$^{-3}$. This indicates that a $\approx 20$~times denser plasma was generated on the wake of the pump beam.

\section*{Discussion}
A possible explanation of these effects is that the dominant physics mechanism arises from the motion of ions triggered by the pump beam~\cite{vieira2012ion,vieira2014ion}.
Similar signatures of an on-axis peak density were theoretically studied in the case of a nonlinear plasma wake~\cite{khudiakov2022ion} and experimentally observed in a Hydrogen plasma with $\alpha>1$ by using shadowgraphy~\cite{gilljohann2019direct}. 
The results were obtained using a pump electron beam while no effect was observed when using a pump laser pulse, indicating that the ion density increase establishes only when triggered by a negatively charged particle beam. This, however, strongly depends on the beam parameters and on the gas type and density. 
As previously discussed, another experiment performed with single ionized Argon showed a depletion of the ion density (resulting in a reduced deceleration of the probe beam) with the initial plasma conditions that were reestablished after $\approx 60$~ns~\cite{d2022recovery}. A similar behavior, attributed to expansion of ions, was observed in a Lithium plasma source again with shadowgraphy~\cite{zgadzaj2020dissipation}. 

\begin{figure}[!ht]
\centering
\subfigure{
\begin{overpic}[width=0.95\linewidth]{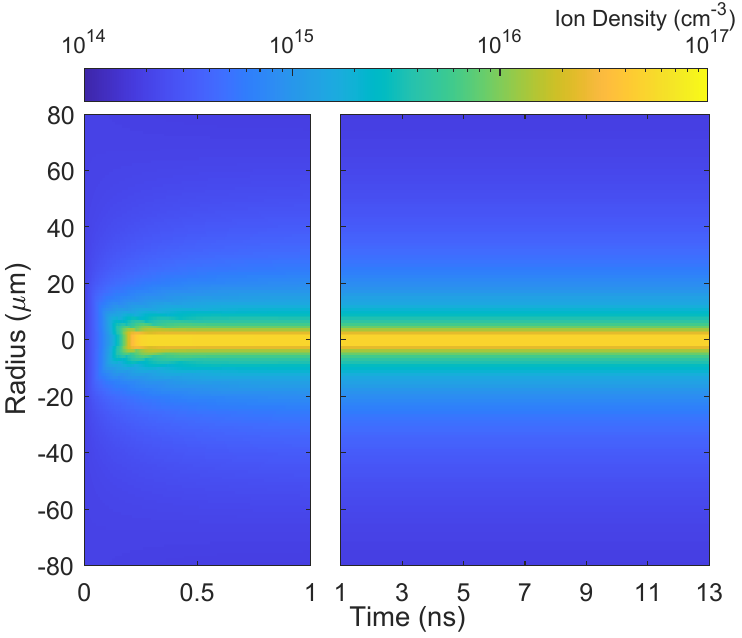}
\put(92,66){\color{white}\textbf{a}}
\end{overpic}
\label{IonDensitySimSP_1ns}
}
\subfigure{
\begin{overpic}[width=1.0\linewidth]{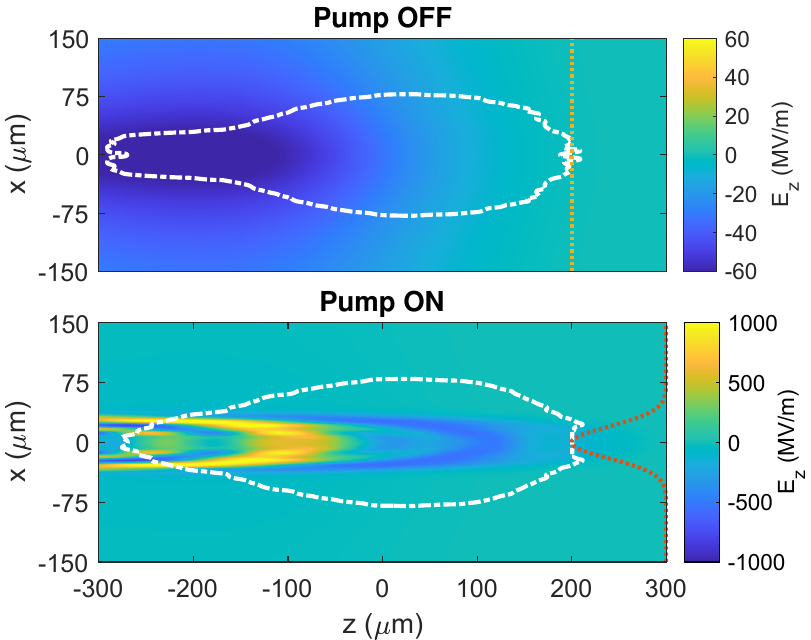}
\put(75,67){\color{black}\textbf{b}}
\end{overpic}
\label{ppWz_tot}
}
\subfigure{
\begin{overpic}[width=1.0\linewidth]{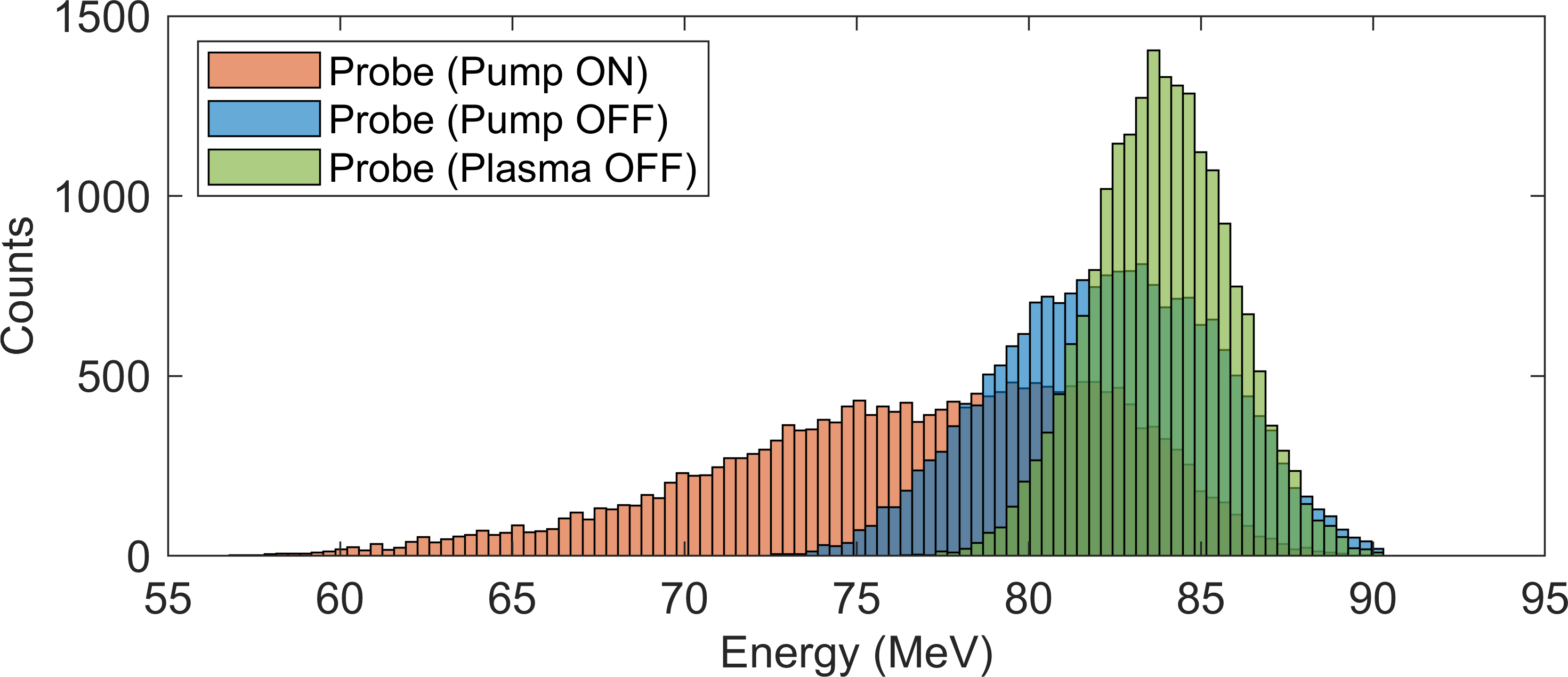}
\put(92,35){\color{black}\textbf{c}}
\end{overpic}
\label{eHist_07ns}
}
\caption{Plasma wakefield simulations. (a) Ion density map obtained after the passage of the pump beam. The initial density was set to $n_p=10^{14}$~cm$^{-3}$. (b) Plasma wakefield acting on the probe beam (whose contour is indicated by the white dashed lines) with the pump beam turned off (top) and on (bottom). The dotted red lines show the radial plasma profiles. (c) Resulting energy distributions of the probe beam. The spectra are obtained at the end of the beam-plasma interaction and over a total length of 5~cm.} 
\label{simData}
\end{figure}

To understand the dynamics of the process we computed the evolution of the ions trajectories after the passage of the pump beam and, in turn, the local ion density as a function of time.
Figure~\ref{IonDensitySimSP_1ns} shows the temporal evolution of the radial ion density computed for an initial plasma density $n_p=10^{14}$~cm$^{-3}$ ($\alpha>1$) from which we can see that the density around the axis becomes several orders of magnitude larger than the initial one. 
The average plasma density over the beam spot size $\sigma_r$ is $n_p\approx 3.5\times 10^{15}$~cm$^{-3}$, in good agreement with the factor 20 we experimentally found.
We can see that the ion density peak survives at least 13~ns, i.e. the maximum delay $\Delta t$ used in the experiment while the timescale over which the on-axis ion-density peak builds up is $\approx 0.2$~ns, thus below the minimum $\Delta t$ we were able to probe.

To provide a comparison with the energy spectra showed in Fig.~\ref{eSpectrum_13ns}, we computed a numerical simulation of the plasma wakefield acting on the probe beam with and without the ion density peak. 
The simulation is performed by means of a numerical code with cylindrical symmetry that tracks the evolution of the beam particles by computing the plasma response~\cite{lu2006nonlinear}. 
Figure~\ref{ppWz_tot} shows the longitudinal wakefield $E_z$ excited in the plasma with pump turned off and on. In the first case the background plasma density is set to $n_p=10^{14}$~cm$^{-3}$, while in the second case the on-axis density peak (induced by the pump beam) is included.
Figure~\ref{eHist_07ns} shows the resulting energy spectra, including the one obtained with the plasma turned off.

\section*{Conclusions}
In conclusion, we performed a proof-of-principle experiment to study the interaction of ultra-short high-density electron beams with a Hydrogen plasma. The experiment demonstrates that for $n_b/n_p<1$ the plasma ions rapidly recover in less than a nanosecond, enabling the operation of the plasma wakefield accelerator at very large repetition rates. 
To increase the accelerating fields and still be able to work at such rates it is thus necessary to raise both plasma and bunch densities while keeping the same density ratio.
This does not consider other effects that may limit the repetition rate such as the heat that is deposited in the capillary walls. This, however, is out of the current study whose aim is to provide an estimate of the recovery time after the beam-plasma interaction.
Ion motion features are evident when $n_b/n_p>1$, where we observed the formation of a locally denser plasma. By measuring the amount of deceleration experienced by the probe beam, we found that the ion density locally increases in a fraction of nanosecond and lasts at least 13~ns.


\section*{Methods}\label{sec:methodsec}
\subsection*{Generation of the pump-probe electron bunches}
The electron bunches are directly generated on the photo-cathode by two ultraviolet laser pulses. Their short durations are obtained with the velocity-bunching technique that requires injection in the first traveling-wave section at the zero crossing of the RF wave. The longitudinal compression is then achieved by slightly accelerating the tail of the beam while decelerating its head. This leads to a rotation of the beam longitudinal phase space that is simultaneously chirped and compressed. To avoid uncontrolled emittance growth, solenoids embedded on the linear accelerator are turned on and provide the necessary extra transverse focusing~\cite{ferrario2010experimental}.
To reach tens of nanoseconds delays between the pump and probe beams, the probe laser is generated by sending a replica of the pump laser over a several meters long delay-line. 
Since the gun operates at $f_{RF}=2.856$~GHz RF frequency (corresponding to an RF period of $T_{RF}\approx 0.35$~ns), the probe laser is injected several RF buckets behind the pump laser but at the same relative RF phase such that the two bunches undergo approximately the same evolution along the photo-injector. The temporal delay $\Delta t$ between the two bunches can therefore be increased in steps of $T_{RF}$.

\subsection*{Plasma source}
For the experiment, a 3~cm-long capillary with a 2~mm-diameter hole has been used. It is made by sapphire and has two inlets for the gas injection. The capillary is installed in a vacuum chamber directly connected with a windowless, three-stage differential pumping system that ensures $10^{-8}$~mbar pressure in the RF linac while flowing the gas. This solution allows to transport the beam without encountering any window, thus not degrading its emittance by multiple scattering~\cite{pompili2017experimental}. The plasma is produced by ionizing the hydrogen gas produced by water electrolysis (Linde NM Plus Hydrogen Generator). A high-speed solenoid valve, located 5~cm from the capillary, is used to fill the capillary with the gas. The valve is opened for 3~ms and the discharge current is applied 1~ms after its closure. The discharge current, applied to the two capillary electrodes, is generated by a high-voltage generator that provides 5~kV pulses with 120~A current through the capillary. The current is monitored with a Pearson current monitor. The plasma density is measured with a Stark-broadening-based diagnostics measuring the H$_{\alpha,\beta}$ Balmer lines. The stabilization of the discharge process and plasma formation is provided by pre-ionizing the gas with a Quantel 532~nm CFR Ultra Nd:YAG laser, reducing the discharge timing-jitter to the order of a few nanoseconds~\cite{articolo_angelo}. The laser is installed close to the capillary vacuum chamber and is injected into it by means of several metallic mirrors.

\subsection*{Modeling of the plasma ion motion}
\begin{figure}[ht]
\centering
\includegraphics[width=1.0\linewidth]{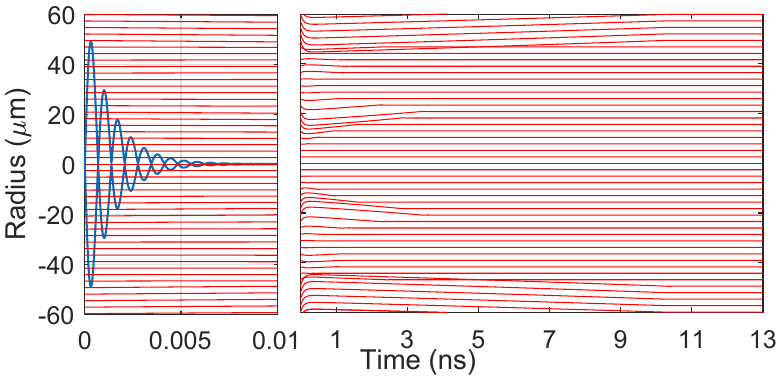}
\caption{Simulation of the plasma response for $n_p\gtrsim n_b$. The initial density was set to $n_p=2\times 10^{16}$~cm$^{-3}$ and the passage of the pump beam is at $t=0$. (b) Trajectories (red) of the ions after the initial attraction operated by the electron beam. The evolution and damping of the blowout region (blue) is also reported.}
\label{IonRadiiPlot_2e16}
\end{figure}

A simplified numerical model to simulate the dynamics of ion motion has been implemented to cross-check the results obtained in the experiment.
Considering that the charge separation in the plasma is governed by the electron beam in a region of radius $r_e$, we assume it is free of plasma electrons as in the blowout regime~\cite{lu2006nonlinear}.
Moreover, having an ultra-relativistic velocity, the beam electric field is mainly concentrated in the transverse direction thus the dynamics evolves only in this plane.
The evolution of the plasma wakefield excited by the pump beam is computed considering that it dissipates due to the presence of collisional friction between the plasma electrons, ions and neutrals~\cite{khodachenko2004collisional}.
The model includes therefore the following equations of motion
\begin{align}
m_e r_e''(t) &+ \beta_e(t) r_e'(t) + k_e(t) r_e(t)=0 \label{eq_e}\\
m_i r_i''(t) &+ \beta_i(t) r_i'(t) + k_i(t) r_i(t) + F_p(t)=0  \label{eq_i}
\end{align}
where $m_{e,i}$ are the electron and ion masses, $r_e$ the radius of the blowout region, $r_i$ the position of the ion, $\beta_{e,i}$ the damping factors due to the frictional forces arising from electrons and ions collisions, the force $F_p=-\nabla p /n_i$ due to pressure $p$ acting on the ion density $n_i$ and $k_{e,i}$ the plasma wakefield term~\cite{ariniello2019transverse}.
The model assumes cylindrical symmetry with the ions to be located at positions $r_i$ and the same is assumed for the blowout region, whose radius $r_e$ is calculated as a single electron oscillation.
The wakefield force term $F=k_{e,i}\cdot r_{e,i}$ is responsible for the oscillations of electrons and ions and is given by
\begin{equation}
k_{i}= -Z\cdot k_e=-{{(Z q_e)^2 n_i}\over{3\epsilon_0}}~,
\label{kfoc_term}
\end{equation}
with $Z$ the ionization degree ($Z=1$ in our case), $q_e$ the electron charge, $\epsilon_0$ the vacuum permittivity and $n_{e,i}$ the electronic and ionic plasma densities. The wakefield is null outside of the blowout region (where we assume the plasma is neutral) and thus $k_i=0$ for ions located at $|r_i|>|r_e|$.
The wakefield damps after several oscillations due to the collisions between the plasma electrons and ions and neutral atoms according to the friction force $F=\beta_{e,i}\cdot r_{e,i}'$, where $\beta_{e,i}=m_{e,i}\cdot f_{e,i}$ with $m_{e,i}$ the electron and ion masses and $f_{e,i}$ the electron and ion collision frequencies. Such frequencies are in turn given by the sum $f_e=f_{ei}+f_{ee}+f_{en}$ and $f_i=f_{ii}+f_{ie}+f_{in}$, i.e. by taking into account the collisions between electrons, ions and neutral atoms~\cite{eliezer2002interaction,zhdanov2002transport}. The collision frequency expressions for the electrons are given by
\begin{align}
f_{ei}&={4\over 3}\sqrt{2\pi \over m_e} n_i Z^2 q_e^4 {\log \Lambda \over{(4\pi \epsilon_0)^2 (k_B T)^{3/2}}}\\
f_{ee}&={f_{ei} \over \sqrt{2}}\\
f_{en}&= n_n \sqrt{8 k_B T\over {\pi m_e}} \Sigma_{en}
\end{align}
where $\log \Lambda$ is the Coulomb logarithm, $k_B$ the Boltzmann constant, $T$ the plasma temperature, $n_n$ the density of the neutral atoms and $\Sigma_{en}\approx 10^{-15}$~cm$^{-2}$~\cite{khodachenko2004collisional}. 
For our experimental conditions we used $T\approx 1$~eV and $n_n\approx 5\times 10^{17}$~cm$^{-3}$.
It is worth to notice that the contribution arising from collisions with neutrals can become rather important, especially during recombination at low plasma densities.
The ion collision frequencies are directly derived from the electron frequencies via~\cite{zhdanov2002transport} as
\begin{align}
f_{ii}&=f_{ei} Z^2 \sqrt{m_e\over {2 m_i}} \\
f_{ie}&=f_{ei}{{n_e m_e}\over{n_i m_i}} \\
f_{in}&= f_{en} {\Sigma_{in} \over \Sigma_{en}}
\end{align}
where $n_e$ is the electron density and and $\Sigma_{in}\approx 5\times 10^{-15}$~cm$^{-2}$. Again, within the blowout region we assume $n_e=0$ while outside of it $n_e=n_i$.
To give an order of magnitude of the expected collision frequencies, for the plasma density range described so far ($n_{i,e}\approx 10^{14-16}$~cm$^{-3}$) the electronic collision rates are $f_{ei}\approx f_{ee}\approx 10^{1-3}$~GHz and $f_{en}\approx 10^{2}$~GHz while the ionic ones are $f_{ii}\approx 1-10$~GHz and $f_{ie}\approx f_{in}\approx 0.1-1$~GHz.
The last force term included in the model is represented by the pressure acting on the ions as $F_p=-\nabla p /n_i$ where the pressure is computed as $p=n_i k_B T$.
The displacement of the plasma ions driven by the pump beam is computed from the beam electric field $E_b$. From the impulse-momentum theorem their initial velocities are given by $v_{i}=Z q_e/m_{i}\int E_b (t) dt$.
Figure~\ref{IonRadiiPlot_2e16} shows the temporal evolution of the ion trajectories when the initial plasma density is set to $n_p=2\times 10^{16}$~cm$^{-3}$. In this case the ions are slightly perturbed by the pump beam, producing negligible effects on the plasma state. The plasma wakefield excited by the beam, as reported, is expected to vanish in less than 10~ps due to the larger viscosity arising from the larger plasma density ($\beta_e\propto f_e\propto n_i$).

\subsection*{Comparison with previous results}
The model has been validated for a wide range of pump-probe delays and plasma densities described in the present work. A cross-check is also provided for the data reported in previous works~\cite{gilljohann2019direct,zgadzaj2020dissipation,d2022recovery} whose findings were attributed to ion motion. The aim of such a comparison is to provide a qualitative interpretation of the ion dynamics involved in the process. In these cases, we used as input parameters the pump beams and plasma configurations employed in such works.

\begin{figure}[ht]
\centering
\includegraphics[width=1.0\linewidth]{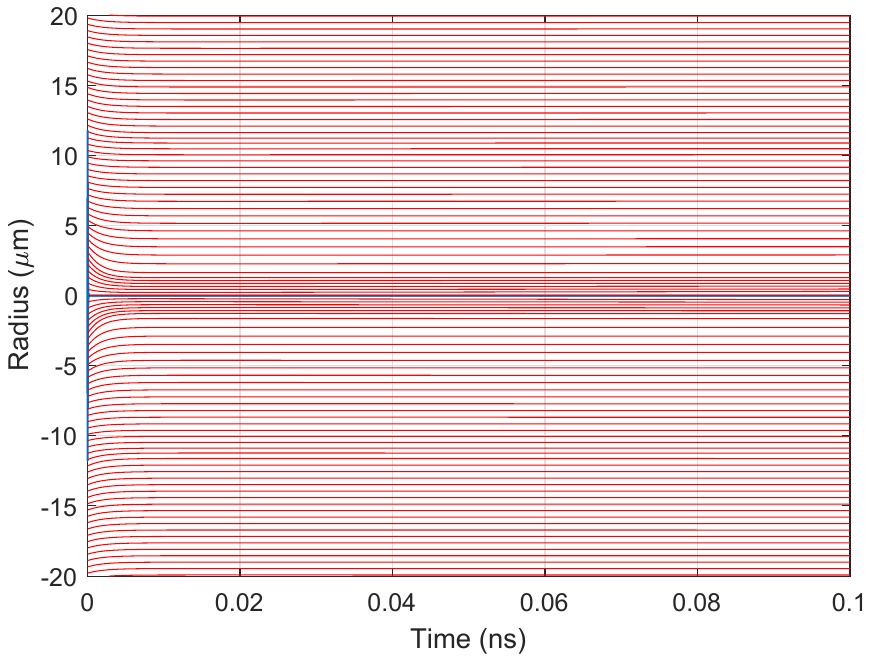}
\caption{Simulation of the plasma results obtained with Hydrogen~\cite{gilljohann2019direct}. Trajectories (red) of the ions after the initial attraction operated by the electron beam (at $t=0$). The initial density is set to $n_p=6\times 10^{18}$~cm$^{-3}$. The beam and plasma parameters are provided in the reference work.}
\label{IonRadiiSim_H_6e18}
\end{figure}

The first experiment was performed using the ATLAS laser at the Laboratory for Extreme Photonics (Garching, Germany) and consisted of a hybrid plasma-wakefield acceleration (PWFA) scheme in which the driver electron beam is generated from a laser-wakefield accelerator (LWFA)~\cite{gilljohann2019direct}. The setup employed two gas-jets with the studies on the plasma waves and ion dynamics carried out on the second jet. The diagnostics consisted of a few-cycle shadowgraphy, well suited to perform studies on high plasma densities. The experiment used a Hydrogen plasma source with density $n_p\approx 6\times 10^{18}$~cm$^{-3}$ and the plasma wakefield was generated using a pump electron beam with 550~pC charge, energy $\approx 100-350$~MeV, $5$~fs bunch duration and $12~\mu m$ transverse spot size.
Figure~\ref{IonRadiiSim_H_6e18} shows the ions trajectories computed by the model for such plasma and beam parameters. Similarly to what is measured in our work, the authors observed the formation of a dense ion channel along the beam path in the same regime ($\alpha>1$). 

\begin{figure}[ht]
\centering
\includegraphics[width=1.0\linewidth]{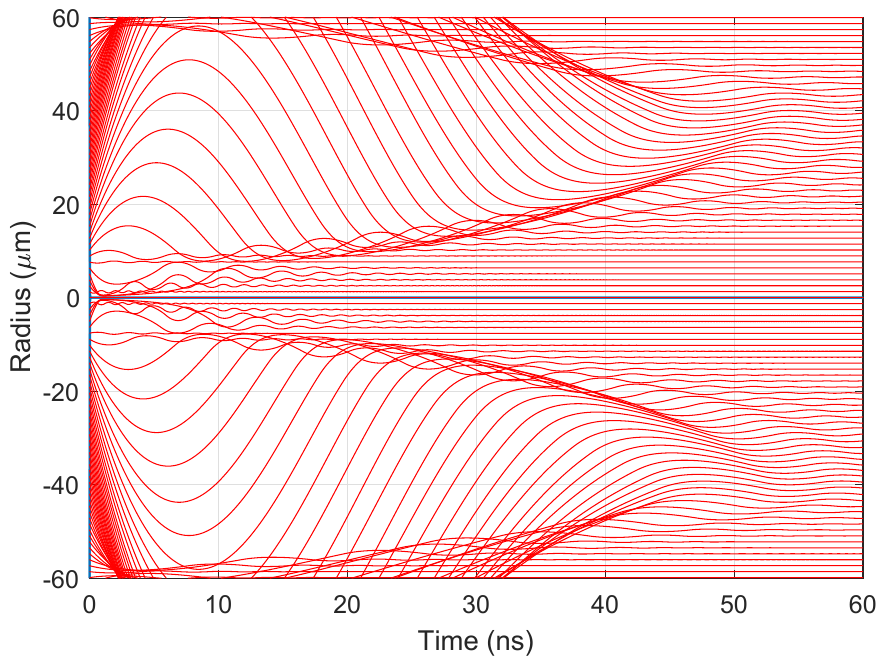}
\caption{Simulation of the plasma results obtained with Argon~\cite{d2022recovery}. Trajectories (red) of the ions after the initial attraction operated by the electron beam (at $t=0$). The initial density is set to $n_p=1.8\times 10^{16}$~cm$^{-3}$. The beam and plasma parameters are provided in the reference work.}
\label{IonRadiiSim_Ar_2e16}
\end{figure}

A second experiment was performed at the FLASHForward facility in DESY (Hamburg, Germany). The experimental setup was very similar to the one described in the current manuscript with two pump and probe electron beams acting as drivers. The authors also used a probe witness bunch (following the driver probe bunch) to characterize the accelerating wakefield in addition to the decelerating one. The plasma source was also similar (plasma confined in a discharge capillary) except for the gas that was employed. The experiment used an Argon plasma with density $n_p\approx 1.8\times 10^{16}$~cm$^{-3}$. The plasma wakefield was generated using a pump electron beam with 590~pC charge, 1.06~GeV energy, $195~\mu m$ length and $5~\mu m$ transverse spot size. Figure~\ref{IonRadiiSim_Ar_2e16} shows the ions trajectories computed by the model for such plasma and beam parameters. It is immediately evident the different signature in the ions trajectories that, depending on their initial locations, are initially pinched by the pump beam and then move outward evacuating the axial region. This produces a smaller deceleration of the probe beam, as observed by the authors. The time needed to completely restore the plasma ions is of the order of 50-60~ns, in agreement with the experimental findings.

\begin{figure}[ht]
\centering
\includegraphics[width=1.0\linewidth]{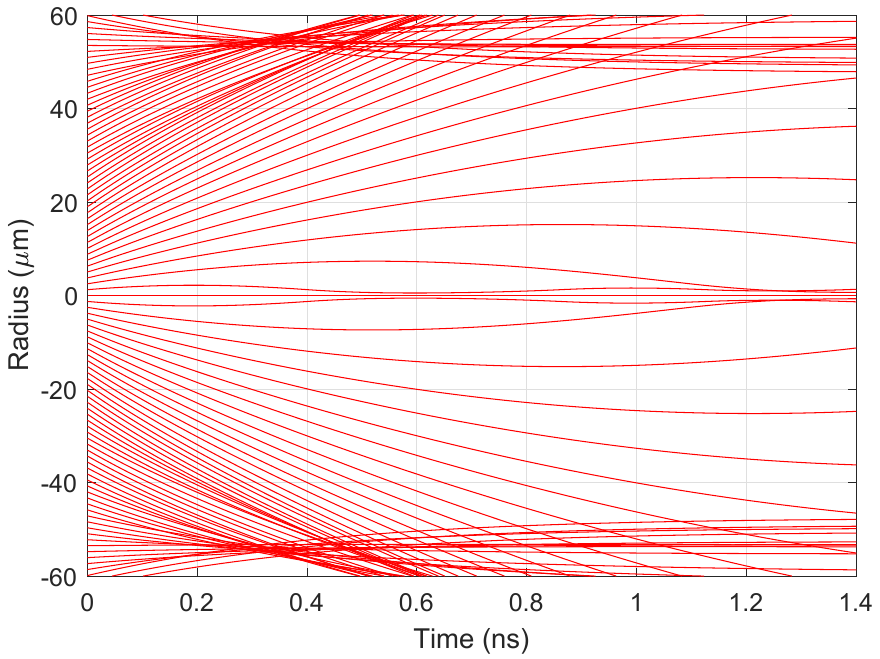}
\caption{Simulation of the plasma results obtained with Lithium~\cite{zgadzaj2020dissipation}. Trajectories (red) of the ions after the initial attraction operated by the electron beam (at $t=0$). The initial density is set to $n_p=8\times 10^{16}$~cm$^{-3}$. The beam and plasma parameters are provided in the reference work.}
\label{IonRadiiSim_Li_8e16}
\end{figure}

A third experiment was performed at the FACET facility at SLAC (Menlo Park, USA) with the goal to monitor the dissipation of the plasma wakes excited by a pump electron beam~\cite{zgadzaj2020dissipation}. The plasma source consisted of Lithium vapor from a heat-pipe oven. The plasma density was set to $n_p\approx 8\times 10^{16}$~cm$^{-3}$ and the wakefield was generated using a pump electron beam with 2~nC charge, 20~GeV energy, $55~\mu m$ length and $30~\mu m$ transverse size.
Figure~\ref{IonRadiiSim_Li_8e16} shows the ions trajectories computed by the model for such plasma and beam parameters. Similar to the previous case, the authors observed the outward motion of the plasma ions after the passage of the pump beam over a timescale of 1.4~ns using shadowgraphy as a diagnostics technique.



\section*{References}
\bibliography{biblio}
\bibliographystyle{apsrev4-2}

\section*{Data Availability}
The data that support the findings of this study are available from the corresponding author on reasonable request.

\section*{Acknowledgments}
This work has has received funding from the European Union's Horizon Europe research and innovation programme under Grant Agreement No. 101079773
(EuPRAXIA Preparatory Phase) and the INFN with the Grant No. GRANT73/PLADIP. We thank G. Grilli and T. De Nardis for the development of the HV discharge pulser, F. Anelli for the technical support and M. Zottola for the experimental chamber installation. We also thank all the
machine operators involved in the experimental run.

\section*{Author contributions}
A.C., M.G. and R.P. planned and managed the experiment, with inputs from all the co-authors. A.B. and L.C. provided the plasma characterization. M.G. and F.V. managed the photo-cathode laser system. V.S. managed the beam diagnostics. J.K.J and T.P. performed laser alignment during the experiment. R.P. carried out the data analysis. R.P. and G.P. provided numerical simulations. V.L. managed the vacuum system. G.D.P. managed the control system. R.P. and A.Z. wrote the manuscript. M.P.A, M.C, E.C., G.C., A.D.D., M.D.G., F.D., M.F., A.G., S.R. and G.J.S. were involved in the experiment, extensively discussed the results and reviewed the manuscript.

\section*{Competing interests}
The authors declare no competing interests.

\end{document}